\documentclass[%
aip,
rsi,%
amsmath,amssymb,
reprint,%
]{revtex4-1}

\usepackage{graphicx}% Include figure files
\usepackage{dcolumn}% Align table columns on decimal point
\usepackage{bm}% bold math
\usepackage[ruled,vlined]{algorithm2e}

\usepackage{xcolor, soul}
\sethlcolor{green}

\begin{document}

\preprint{AIP}

\title[]{Polyatomic radiative association by quasiclassical trajectory calculations: \\ Cross section for the formation of HCN molecules in H + CN collisions}% Force line breaks with \\
%\thanks{Footnote to title of article.}

\author{P\'eter Szab\'o}
\email{peter88szabo@gmail.com}
% \affiliation{Department of Physics and Material Science, University of Luxembourg, Luxembourg}%
\affiliation{Applied Physics, Division of Materials Science, Department of Engineering Science and Mathematics,
Lule\aa \space University of Technology, 97187 Lule\aa, Sweden}

\author{Magnus Gustafsson}%
\email{magnus.gustafsson@ltu.se}
\affiliation{Applied Physics, Division of Materials Science, Department of Engineering Science and Mathematics,
Lule\aa \space University of Technology, 97187 Lule\aa, Sweden}

% \homepage{http://www.Second.institution.edu/~Charlie.Author.}
%\affiliation{%
%5Second institution and/or address%\\This line break forced% with \\
%5}%

\date{\today}% It is always \today, today,

\begin{abstract}
We have developed the polyatomic extension of a recently established (M. Gustafsson, J. Chem. Phys, 138, 074308 (2013)) classical theory of radiative association in the absence of electronic transitions. The cross section of the process is calculated by a quasiclassical trajectory method combined with the classical Larmor formula which can provide the radiated power in collisions. We have also proposed a double-layered Monte Carlo scheme for efficient computation of ro-vibrationally quantum state resolved cross sections for radiative association. Besides the method development, we also calculated and fitted the global potential energy and dipole surface for the H + CN collisions to test our polyatomic semiclassical method.
\end{abstract}

%\keywords{radiative association, surface-hopping, trajectory, interstellar, spontaneous emmision}%Use showkeys class option if keyword
\maketitle

%\begin{quotation}
%The ``lead paragraph'' is encapsulated with the \LaTeX\ 
%\verb+quotation+ environment and is formatted as a single paragraph before the first section heading. 
%(The \verb+quotation+ environment reverts to its usual meaning after the first sectioning command.) 
%Note that numbered references are allowed in the lead paragraph.
%%
%The lead paragraph will only be found in an article being prepared for the journal \textit{Chaos}.
%\end{quotation}

%============================================================================================================
\section{\label{sec:level1}Introduction}

In astronomical environments there are several physical and chemical processes which result in molecule production: i) chemical exchange reactions in gas phase or on the surface of grains, ii) collision with electrons and iii) interaction of atoms and molecules with radiation. In complex-forming bimolecular collisions the transient complex can be stabilized either by the collision of a third body or by emitting a photon. The latter process is called radiative association (RA) and it contributes to molecule formation in dust-poor regions of interstellar space \citep{Babb_Kirby_98, Rev_theo}. 
%In general, the rate coefficient of RA is much smaller than the rate of exchange reactions. In spite of this, RA can be considerable in the low-density region of the astrochemical environment since it is a direct pathway to molecule formation, in contrast to tangled chemical network where multiple consecutive reactions,  often including polyatomic reactants,  are needed for the formation of given chemical species. Besides the contribution to the molecule formation, radiative association and radiative quenching (i.e. transitions to the continuum) also serves as an energy dissipation channel. That is why these collision induced processes might also have a role in the cooling process of molecular nebulae which is necessary for star formation. 

Owing to the very small probability of spontaneous photon emission, it is difficult to measure the cross section and rate constant of RA. Experimental studies have so far been carried out only for a few ionic systems \cite{Gerlich_92_cr_92_1509}. That is why theoretical modeling of the dynamics of RA is the feasible way to obtain a reliable cross section or rate constant of the process  \cite{Rev_theo}. 
Nevertheless, RA is also challenging to study theoretically since, in principle, it requires the quantum mechanical description of the time-dependent process. %especially when the stabilization of the collision-complex takes place through electronic transitions.
Due to the difficulty of quantum mechanical treatments, most of the previous dynamical calculations have focused merely on diatomic molecule formation \citep{Radas_1,Radas_2,Radas_3,Radas_4,Radas_5,Radas_6,Radas_7,Radas_8,Radas_9,Radas_10,Radas_11,Radas_12,Radas_13,Radas_14,Radas_15,Radas_16, Radas_17, Radas_18,Radas_19,Radas_20,Radas_21,Radas_22,Radas_23,Radas_24,Radas_25,Radas_26,Nonadia, LTE_theory}.
% Radas_7 och Radas_19 is the same paper. I removed 7.
%Peter: Now I changed number of references, hence, Radas_7 is what was originally Radas_8, and so on with the other refernces (the indices are shifted with one)
There are only a few quantum dynamics studies where radiative association of triatomic molecules has been considered in full dimension including the following molecules \citep{Polyatom_1,Polyatom_2,Polyatom_3,Polyatom_4,Polyatom_5, Polyatom_6, Polyatom_7, Polyatom_8}: HeH$_{2}^{+}$, H$_{3}^{-}$ , HN$_{2}^{-}$, AlH$_{2}^{+}$, AlD$_{2}^{+}$, NaH$_{2}^{+}$, NaD$_{2}^{+}$, HCO$^{-}$, HCO.
% Include references to the publications with those studies of triatomics.
For other polyatomic cases, either reduced dimensional semiclassical dynamical calculations \citep{Polyatom_RedDim_1,Polyatom_RedDim_2} or statistical rate theories\citep{Polyatom_Stat_1,Polyatom_Stat_2,Polyatom_Stat_3, Polyatom_Stat_4} are applied to obtain their rate constant.

The extension of the theoretical
models for polyatomic systems is desirable, since the bigger the reactants, the more probable the RA is \cite{Polyatom_Stat_1}. Unfortunately, the quantum dynamical methods are not feasible for larger polyatomic molecules. However, the classical or semiclassical treatment of the dynamics can provide a practical but still efficient solution regardless the size of reactants. Despite the lacking quantum effects, the trajectory based method with a good quality potential energy surface can provide accurate rate constants, spectral properties and also reliable atomic-scale mechanism of molecular collisions \citep{QCT_1,QCT_2,QCT_3,QCT_4,QCT_5,QCT_6, QCT_7, QCT_8}.

\begin{figure}
\includegraphics[width=8.5cm,angle=0]{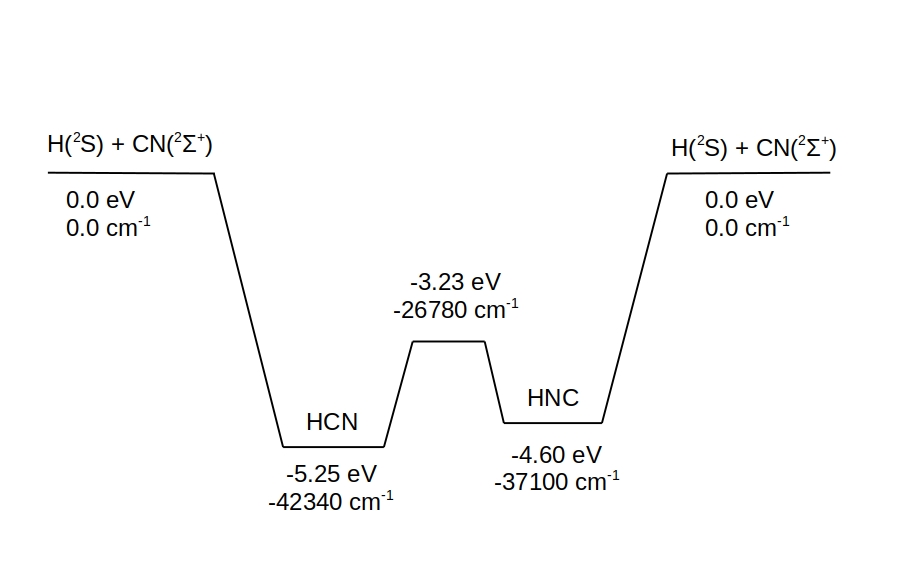}
\caption{Schematic representation of H + CN potential energy surface}
\label{fig:PESscheme}
\end{figure}

Two trajectory-based classical methods have been recently established for RA. One is for RA in absence of electronic transitions  based on the classical Larmor theory of radiation \cite{Classical},
% cite Gustafsson 2013.
and another for the calculation of RA through electronic transitions based on the semiclassical surface-hopping formalism \cite{SurfaceHopping}.
% cite Szabó and Gustafsson 2017.. Both methods are tested and used successfully to calculate the diatomic molecule formation through RA. However, these methods have not yet been extended for polyatomic systems to consider the dynamics in full dimension.

The purpose of this work was to develop and implement the polyatomic extension of the earlier established method for RA in absence of electronic transitions \cite{Classical}. In order to test our method we calculated the cross sections  of the following reaction: 
\begin{equation}
\text{H}\left(^{2}S\right) + \text{CN}\left(^{2}\Sigma^{+}\right) \rightarrow  \text{HCN/HNC} + \hbar \omega
\label{reaction}
\end{equation}
Besides the dynamical studies, we also developed a global ab-initio potential energy (see Fig~\ref{fig:PESscheme}.) and dipole surface for H + CN collisions.

%================================================================================

\section{\label{sec:level1b} Quasiclassical theory of radiative association}

According to the recently developed method~\citep{Classical}, the cross section of the radiative association can be computed by using a combination of the classical trajectory method and the Larmor formula in which the spontaneous emission is a result of a time-dependent dipole.  The  method  applies only in absence of non-electronic  transitions, and it was tested so far only for diatomic molecule formation. As is presented earlier, this  classical method can reproduce the resonance free quantum mechanical cross sections  with a good accuracy. Here, we generalize this classical theory of radiative association~\citep{Classical} for arbitrary polyatomic system by deriving the formula for quantum state resolved semiclassical cross section. Then we also give two different Monte Carlo schemes for the efficient evaluation of the cross section.
Note that in this work the term {\it semiclassical} refers to the semiclassical quantization of the rovibrational states of the reactants. This should not be confused with the semiclassical treatment of radiative association of diatomic molecules involving different electronic states\citep[]{Bates_51_MNRAS_111_303,Zygelman_88_PhysRevA_38_1877}.

\subsection{Derivation of cross section formula}
Since radiative association is a collision process, its cross section can be obtained by the classical collision theory
\begin{equation}
\sigma\left(E_{c}\right)=2\pi f_{\text{stat}} \intop_{0}^{\infty}b \, P_{\text{r}}\left(E_{c},b\right)\text{d}b ,
\label{class_cross}
\end{equation}
where $P_{\text{r}}$ is the opacity function (the probability of the radiation as a function of impact parameter, $b$, and collision energy, $E_{c}$). The statistical weight factor, $f_{\text{stat}}$ can be calculated from the spin- and orbital angular momentum multiplicity of the electronic quantum states of the reactants and product involved in the collision. 

In the frequency interval $\left[\omega,\omega+\text{d}\omega\right]$ the probability of emission is $I\left(\omega\right)\text{d}\omega/\hbar\omega$. Hence, the total probability of radiation is
\begin{equation}
P_{\text{r}}=\intop_{0}^{\infty}\frac{I\left(\omega\right)}{\hbar\omega}\text{d}\omega,
\label{Pr_gen}
\end{equation}
where $I\left(\omega\right)$ is called radiative power or the intensity of emission. In the classical theory of dipole radiation, the time-dependent radiative power (the total energy radiated from the system per unit time) is given by the Larmor formula~\citep{LandauLifsitz}
\begin{equation}
I(t)=-\frac{\text{d}E}{\text{d}t}=\frac{2}{3c^{3} \, 4\pi\epsilon_{0}} \mathbf{\ddot{D}}^{2},
\label{Larmor_t}
\end{equation}
where $E$ is the radiated energy, $c$ is the speed of light and $\mathbf{D}$ is time-dependent dipole moment of the system. By Fourier transforming Eq.~(\ref{Larmor_t}) and using the Parseval theorem
\begin{equation}
\intop_{-\infty}^{\infty}f_{t}^{2}\text{d}t=\intop_{-\infty}^{\infty}\left|f_{\omega}\right|^{2}\frac{\text{d}\omega}{2\pi}=2\intop_{0}^{\infty}\left|f_{\omega}\right|^{2}\frac{\text{d}\omega}{2\pi},
\label{Parseval}
\end{equation}
we obtain the frequency dependent radiated power as~\citep{LandauLifsitz}
\begin{equation}
I(\omega)=\frac{2}{3c^{3} \pi \, 4\pi\epsilon_{0}} \left| \intop_{-\infty}^{\infty}e^{i\omega t} \mathbf{\ddot{D}}\left(b,t,E_{c}\right)\text{d}t\right|^{2}.
\label{Larmor_w}
\end{equation}
Inserting $I(\omega)$ into Eq.~(\ref{Pr_gen}), the total radiative probability can be computed as
\begin{equation}
P_{\text{r}} = \frac{2}{3c^{3} \pi \hbar \, 4\pi\epsilon_{0}}     \intop_{\omega_{c}}^{\omega_{\text{max}}} \frac{1}{\omega}\left| \intop_{-\infty}^{\infty}e^{i\omega t} \mathbf{\ddot{D}}\left(b,t,E_{c}\right)\text{d}t\right|^{2}\text{d}\omega,
\label{Pr_gen_simple}
\end{equation}
where the new boundaries of the frequency integral are introduced to take into account only that transitions which result in stable molecule formation. By having $\omega_c = E_c/\hbar$ as lower limit, rather than 0, transitions to the continuum are excluded. Furthermore, $\omega_{\text{max}}=E_{\text{max}}/\hbar$, where $E_{\text{max}}$ is the maximally possible energy loss of the system in the collision corresponding to the minimum possible ro-vibrational energy and total angular momentum of the formed molecule. See the Appendix~(\ref{sec: App-Emax}) for further details of how to determine the value of $E_{\text{max}}$.
For neutral systems we may introduce further simplification by applying the differentiating property of Fourier transformation
\begin{equation}
|\mathcal{F} (\mathbf{\ddot{D}}) |^{2}=\omega^{4}\left|\mathcal{F}\left(\mathbf{D}\right)\right|^{2},
\label{derivative}
\end{equation}
with this identity 
\begin{equation}
P_{\text{r}} = \frac{2}{3c^{3} \pi \hbar  4\pi\epsilon_{0}}     \intop_{\omega_{c}}^{\omega_{\text{max}}}\omega^{3}\left| \intop_{-\infty}^{\infty}e^{i\omega t} \mathbf{D}\left(b,t,E_{c}\right)\text{d}t\right|^{2}\text{d}\omega.
\label{Pr_neut_simple}
\end{equation}
It should be stressed that Eq.~(\ref{Pr_neut_simple}) is valid solely for neutral species. For ionic system $P_r$ has to be calculated from Eq.~(\ref{Pr_gen_simple})  (see e.g. \"Ostr\"om et al. \cite{Radas_19}). In such cases the identity Eq.~(\ref{derivative}) does not hold because the dipole moment of ionic systems is not an $L^2$ integrable function. 

Eqs.~(\ref{Pr_gen_simple}) and (\ref{Pr_neut_simple}) are suitable to describe merely the collision of two structureless participles. In polyatomic reactions the probability of the process also depends on the ro-vibrational states of the interacting partners. In the semiclassical picture these are characterized by a collection of initial (semiclassical) ro-vibrational quantum numbers ($\mathbf{n}_{AB}$) and the phases of rotations ($\boldsymbol{\eta}$) and vibrations ($\boldsymbol{\chi}$) of the reactants. Furthermore, the relative orientation of the colliding fragments, which is characterized by the Euler angles ($\boldsymbol{\psi}$), are also needed to be considered. Hence, in general case (either ionic or neutral system), the polyatomic radiative probability density takes the form
%%%%%%%%%%%% WIDETEXT%%%%%%%%%%%%%%%%%%%%%%%%%%%%%%%%%%%%%%%%%%%%%%
\begin{widetext}
\begin{equation}
P^{\text{gen}}_{\text{r}}\left(E_{c},b,\mathbf{n}_{AB},\boldsymbol{\eta},\boldsymbol{\chi},\boldsymbol{\psi}\right) = \frac{2}{3c^{3} \pi \hbar \, 4\pi\epsilon_{0}}\intop_{\omega_{\text{c}}}^{\omega_{\text{max}}}\frac{1}{\omega}\left|\intop_{-\infty}^{\infty}e^{i\omega t}\mathbf{\ddot{D}}\left(b,t,E_{c},\mathbf{n}_{AB},\boldsymbol{\eta},\boldsymbol{\chi},\boldsymbol{\psi}\right)\text{d}t\right|^{2}\text{d}\omega.
\label{Pr_gen_long}
\end{equation}
For neutral systems by using Eq.~(\ref{derivative}), Eq.~(\ref{Pr_gen_long}) is simplified 
\begin{equation}
P^{\text{neut}}_{\text{r}}\left(E_{c},b,\mathbf{n}_{AB},\boldsymbol{\eta},\boldsymbol{\chi},\boldsymbol{\psi}\right) = \frac{2}{3c^{3} \pi \hbar \, 4\pi\epsilon_{0}}\intop_{\omega_{\text{c}}}^{\omega_{\text{max}}}\omega^{3}\left|\intop_{-\infty}^{\infty}e^{i\omega t}\mathbf{D}\left(b,t,E_{c},\mathbf{n}_{AB},\boldsymbol{\eta},\boldsymbol{\chi},\boldsymbol{\psi}\right)\text{d}t\right|^{2}\text{d}\omega
\label{Pr_neut_long}
\end{equation}
Inserting Eqs.~(\ref{Pr_gen_long}) and (\ref{Pr_neut_long}) into Eq.~(\ref{class_cross}) and integrating over the phase variables $\left( \boldsymbol{\eta},\boldsymbol{\chi},\boldsymbol{\psi}\right)$, we obtain the initial quantum state resolved semiclassical radiative association cross section for general and neutral polyatomic systems as 
\begin{equation}
\sigma_{\text{gen}} \left(E_{c},\mathbf{n}_{AB}\right) =  \frac{4 f_{\text{stat}}}{3c^{3} \hbar \, 4\pi\epsilon_{0}} \intop_{0}^{\infty}b \, \text{d}b \intop_{0}^{2\pi}\cdots\intop_{0}^{2\pi} \text{d}\boldsymbol{\eta}\,\text{d}\boldsymbol{\chi}\,\text{d}\boldsymbol{\psi} \intop_{\omega_{\text{c}}}^{\omega_{\text{max}}} \frac{1}{\omega} \left|\intop_{-\infty}^{\infty}e^{i\omega t}\mathbf{\ddot{D}}\left(b,t,E_{c},\mathbf{n}_{AB},\boldsymbol{\eta},\boldsymbol{\chi},\boldsymbol{\psi}\right)\text{d}t\right|^{2}\text{d}\omega,
\label{cross_gen_long}
\end{equation}
\begin{equation}
\sigma_{\text{neut}} \left(E_{c},\mathbf{n}_{AB}\right) =  \frac{4 f_{\text{stat}}}{3c^{3} \hbar \, 4\pi\epsilon_{0}} \intop_{0}^{\infty}b \, \text{d}b \intop_{0}^{2\pi}\cdots\intop_{0}^{2\pi} \text{d}\boldsymbol{\eta}\,\text{d}\boldsymbol{\chi}\,\text{d}\boldsymbol{\psi} \intop_{\omega_{\text{c}}}^{\omega_{\text{max}}}\omega^{3}\left|\intop_{-\infty}^{\infty}e^{i\omega t}\mathbf{D}\left(b,t,E_{c},\mathbf{n}_{AB},\boldsymbol{\eta},\boldsymbol{\chi},\boldsymbol{\psi}\right)\text{d}t\right|^{2}\text{d}\omega
\label{cross_neut_long}
\end{equation}
One needs only to follow the time-dependent dipole moment of the system along its classical trajectory to calculate the ro-vibrationally state resolved probability of spontaneous emission (Eq.~(\ref{Pr_gen_long}-\ref{Pr_neut_long})) or the cross section (Eq.~(\ref{cross_gen_long}-\ref{cross_neut_long})) of molecule formation in collisions.
\end{widetext}

%%%%%%%%%%%% WIDETEXT%%%%%%%%%%%%%%%%%%%%%%%%%%%%%%%%%%%%%%%%%%%%%%

\subsection{\label{sec:level2} Monte Carlo estimation of the cross section}
Since the radiative probability as well as the cross section are many dimensional integrals Monte Carlo (MC) integration is needed for the efficient evaluation of Eqs.~(\ref{Pr_gen_long}~-~\ref{cross_neut_long}). To use the MC method one needs to sample randomly the initial conditions of reactants, including the ro-vibrational phases $\left( \boldsymbol{\eta},\boldsymbol{\chi}\right) \in\left[0,2\pi\right]$, the Euler angles $\left( \boldsymbol{\psi}\right)$ and the impact parameter ($b$), to take into account all possible phase space points of colliding fragments that may have a contribution to the reaction.

\subsubsection{Method I: continuous radiation}
When the phase variables are sampled uniformly and the impact parameter is sampled as $b_i=b_{\text{max}}\sqrt{\xi_i}$, where $\xi_{i} \in\left[0,1\right]$ is a uniform random number, then we obtain the following Monte Carlo formula
\begin{equation}
\sigma \left(E_{c},\mathbf{n}_{AB}\right) \approx \frac{\pi b_{\text{max}}^{2}}{N_{\text{tot}}} \sum_{i}^{N_{\text{tot}}} P_{\text{r}}^{(i)}\left(E_{c},\mathbf{n}_{AB},b_{i},\boldsymbol{\eta}_i,\boldsymbol{\chi}_i,\boldsymbol{\psi}_i\right)
\label{naiv_MC}
\end{equation}
for the initial quantum state resolved cross section for the collision of two arbitrary molecules, where $P_{\text{r}}^{(i)}$ has to be calculated from Eq.~(\ref{Pr_gen_long}) or Eq.~(\ref{Pr_neut_long}) for the $i$th trajectory. 
In this MC scheme each trajectory has a contribution to the total radiative probability.
As we will see in the Computation Detail section, the applicability of Eq.~(\ref{naiv_MC}) breaks down at low collision energies (which is essential in modelling of astrochemical environment) thanks to trajectories with extra long lifetime.

\subsubsection{Method II: discrete photons}
To overcome the shortcomings of Method I., we introduce a double-layered Monte Carlo method. The random sampling of the initial conditions is the same as we did it in the Method I. However, in the second MC layer we introduce another random sampling that is mimicking the quantized nature of the radiation similarly to our recently established surface-hopping formalism\citep{SurfaceHopping} that is used for the description of radiative association with electronic transitions. In this second MC sampling we decide randomly about the fate of trajectories whether happened photon emission during a collision or not. Owing to this, not all the trajectories will have a contribution to the cross section as in Method I. Based on this classification, we can simplify the MC estimation formula of the cross section by counting the number of reactive trajectories
\begin{equation}
\sigma \left(E_{c},\mathbf{n}_{AB}\right) \approx \pi b_{\text{max}}^{2} \frac{N_{\text{rad}}}{N_{\text{tot}} } ,
\label{improved_MC}
\end{equation}
where the number of trajectories where photon emission happened can be calculated as 
\begin{equation}
N_{{\text{rad}}} = \sum_{i}^{N_{{\text{tot}}}} \Theta_{\text{i}} \left(E_{c},\mathbf{n}_{AB},b_{i},\boldsymbol{\eta}_i,\boldsymbol{\chi}_i,\boldsymbol{\psi}_i\right),
\label{Nrad}
\end{equation}
where the characteristic function is defined as
\begin{equation}
\Theta_{i}=\left\{ \begin{array}{ccc}
1, & \text{if} & P_{\text{r}}^{(i)}>\xi_{i}\\
\\
0, & \text{if} & P_{\text{r}}^{(i)}<\xi_{i}
\end{array}\right.
\label{charfunc}
\end{equation}
Eq.~(\ref{charfunc}) means that we are using the total radiative probability of a trajectory to decide randomly about its outcome. The obtained new formula in its present form does not give an advantage.
Eq.~(\ref{improved_MC}) is still inefficient since the probability of the spontaneous emission ($P_r$) is very small in general. This means one needs a huge number of trajectories $(N_{\text{tot}} > 10^8-10^{10})$ to obtain a reliable estimation of the cross section.

To improve the efficiency of the MC formula we are adapting the MC procedure developed originally for the surface-hopping method for radiative association\citep{SurfaceHopping}. To this end we use the fact that the classical Larmor formula can be considered as the classical limit of the Fermi's golden rule (FGR) \citep{Pafutti}. However, in contrast to the Larmor formula the stimulated emission is also naturally involved in FGR not only the spontaneous emission \citep{Davydov,SchatzRatner}. Based on this analogy with the FGR, we introduce a new characteristic function 
\begin{equation}
\Theta_{i}=\left\{ \begin{array}{ccc}
1, & \text{if} & (\Omega + 1) P_{\text{r}}^{(i)}>\xi_{i}\\
\\
0, & \text{if} & (\Omega + 1) P_{\text{r}}^{(i)}<\xi_{i} \, ,
\end{array}\right. 
\label{improved_Charfunc}
\end{equation}
which describes stimulated emission characterized by the number of photons ($\Omega$) in the radiation field. However, to use Eq.~(\ref{improved_Charfunc}) for the characterization of spontaneous emission one needs to scale down Eq.~(\ref{improved_MC}) by the applied photon number 
\begin{equation}
\sigma \left(E_{c},\mathbf{n}_{AB}\right) \approx \frac{\pi b_{\text{max}}^{2}}{\Omega + 1} \frac{N_{\text{rad}}}{N_{\text{tot}} } .
\label{Improved_MC_cross}
\end{equation}
Eqs.~(\ref{improved_MC}) and (\ref{Improved_MC_cross}) result in the same cross section when $N_{\text{tot}} \rightarrow \infty$. However, in practice much less trajectory is needed for proper MC estimation of the cross section by using Eq.~(\ref{Improved_MC_cross}).

The scheme above was defined to characterize spontaneous emission in bimolecular collisions. However, it can be easily modified to compute semiclassical stimulated emission in the presence of strong radiation fields, where $\Omega >> 0$.  In the case of stimulated emission one needs to use Eqs~(\ref{improved_MC}) and (\ref{Nrad}) with the characteristic function defined by Eq~(\ref{improved_Charfunc}) to Monte Carlo estimate the cross section.

\section{\label{sec:level3} Computational details of the dynamical calculations}
\subsection{Details of the quasiclassical trajectory calculations}
The molecular collisions were studied using the QCT method. The calculations were performed using an in-house built code. In QCT calculations, the classical equations of motion are solved for each atom participating in the collision. To simulate the quantized nature of the vibration and rotation of the reactant molecules, the internal motion of molecules is described by ensembles of classical states that correspond to preselected quantum mechanical states. The rotational angular momentum  of the CN molecule, $L$,  was calculated from the conventional semiclassical quantization formula $L=\hbar\sqrt{(j(j+ 1)}$, where $j$ is the rotational quantum number of the diatomic molecule. The ro-vibrational initial state of the CN molecule is sampled by employing a rotating Morse oscillator based on Porter-Raff-Miller method\citep{RotMorse}. The impact parameter $b$ was sampled with a weight proportional to $b$ itself, and the maximum impact parameter was chosen dynamically as a function of the collision energy to take into account all reactive events.  The Hamilton's equations of motion are integrated in Cartesian coordinates using the 10th order Adams-Moulton predictor-corrector algorithm initiated by Velocity-Verlet integrator. The integration time step was 0.5 atomic time unit in the calculations, guaranteeing energy conservation better than 0.5 $\text{cm}^{-1}$ at the given collision energy. In this work, we have calculated $5 \times 10^4$ trajectories at each collision energy.

\subsection{Details for the computation of radiative association cross section}
In this section we discuss the numerical issues and their solutions regarding the computation of the cross section. Along the trajectories we recorded the time-dependent dipole of the system, that provides the starting point for the computation of the cross section. First we discuss how to obtain efficiently the radiative probability for each trajectory and then how to use these individual probabilities to obtain the cross section.

\subsubsection{Which dipole is supposed to be used and how?} \label{sec:WhichDipole}
In general, the non-interacting reactants possess a permanent dipole moment, such as that of CN in our case. This means that rotating and vibrating molecules also radiate classically even before approaching the interaction zone in the collision. This effect should be removed from the recorded $\mathbf{\widetilde{D}}(t)$ because it does not result in molecule formation.
Based on this consideration, the following modified dipole is used in Eq.~(\ref{Pr_gen_long})
\begin{equation}
\mathbf{D}\left(\mathbf{q},t\right)=\left\{ \begin{array}{ccc}
\mathbf{\widetilde{D}}_{\text{eq}}  & \:\text{if}\: & \mathbf{q}\notin\left[\text{int. zone}\right] \, \\
\\
\mathbf{\widetilde{D}}\left(\mathbf{q},t\right) & \:\text{if}\: & \mathbf{q}\in\left[\text{int. zone}\right],
\end{array}\right.
\label{interactionzone}
\end{equation}
where $\mathbf{\widetilde{D}}_{\text{eq}}$ is the constant value of the dipole of the non-interacting fragments at their equilibrium structure. In the case of HCN/HNC system, this is illustrated by non-zero value of $D_z$ in Fig.~\ref{fig:Dip_2D}a, when the H atom is far away, corresponding to the dipole of the free CN molecule.
In other words, outside the interaction zone $\mathbf{D}\left(\mathbf{q},t\right)$ has direction and magnitude that are independent of the CN orientation and bond length, respectively.
Hence, outside the interaction zone, the second derivative of the dipole in Eq.~(\ref{Pr_gen_long}) results in a zero contribution to the radiative power. In this work the interaction zone is chosen based on the Jacobi distance, where $R < 5.0$ \AA. Between  4.5 \AA~and 5.5 \AA~we used a switching function in order to obtain a smooth dipole.
Furthemore, in Appendix~(\ref{sec:App_Dip}) we give an alternative procedure to eliminate the radiation of the non-interaction reactants. In the present work, we tested both methods, and they resulted in the same cross section.

Besides the vibrational contribution of the radiation, it is also important to treat properly the rotational effect. In our work, the dipole vector was written in Cartesian coordinates. Owing to this, the effect of rotation in the radiation process is automatically included. However, it should be stressed that the rotational contribution is not considered when the components of $\mathbf{D}(t)$ are expressed by internal coordinates. In such cases one needs to use also the Euler-angles to take into account the change of dipole stemming from the time-dependent orientation of the collision-complex.

There is a further technical issue when the Cartesian component of dipole is given in the frame of  principal  axis  of  inertia (PAI) -- like the HCN/HCN system in the present work -- and the dynamics is calculated in laboratory fixed Cartesian coordinates (LAB). In such cases the dipole vector is supposed to be transformed from PAI to LAB coordinates in every time-step of the trajectory by using the orthogonal transformation (rotation) matrix formed by the eigenvectors of inertia tensor. However, the sign of the eigenvectors is not defined by the diagonalization which means there are eight different transformation matrices by considering the possible sign combinations. Such ambiguity may introduce discontinuity into the dipole in the LAB frame that is supposed to be used in Eqs.~(\ref{Pr_gen_long}-\ref{Pr_neut_long}). A similar problem appears in reactive dynamics calculation, owing to the sign-ambiguity of the rotation matrix, when the Eckart transformation is utilized to project out the rotational modes~\cite{Czako}. In order to avoid this numerical problem, we transformed the PAI dipole vector into LAB dipole vector by using all of the eight rotation matrices, and at the end we choose that matrix where the obtained LAB dipole shows the largest overlap with the LAB dipole of the previous time-step. Since the inertia tensor has to be diagonalized in each time-step of dynamics we employed the special algorithm of Kopp designed for $3\times3$ matrices~\citep{3x3diag}.

\subsubsection{Evaluation of the radiative probabilities}
For the calculation of the cross section of neutral systems like the HCN molecule it would be rational to use Eq.~(\ref{cross_neut_long}) with Eq.~(\ref{Pr_neut_long}). However, owing to the $\omega^3$ factor in the frequency integral of Eq.~(\ref{Pr_neut_long}), it might appear severe numerical issues when the Fourier transformation of the dipole is noisy. In our numerical experiments, the $\omega^3$ factor amplified the numerical noise of the signal at large frequencies, which resulted in a no-convergent frequency integral for some trajectories. That is why Eq.~(\ref{cross_gen_long}) is recommended even for the characterization of a neutral system instead of  Eq.~(\ref{cross_neut_long}) because it always provides a convergent frequency integral thanks to the $1/\omega$ factor. However, using Eq.~(\ref{cross_gen_long}) demands the calculation of the second time derivative of dipole which may introduce additional noise into the spectrum. To reduce the numerical noise of the signal before Fourier transformation, we applied the Savitzky-Golay filter\cite{SavGol} on $\mathbf{D}(t)$ that automatically can provide the desired second derivative, $\mathbf{\ddot{D}}(t)$.
\begin{figure}
\includegraphics[width=8.5cm,angle=0]{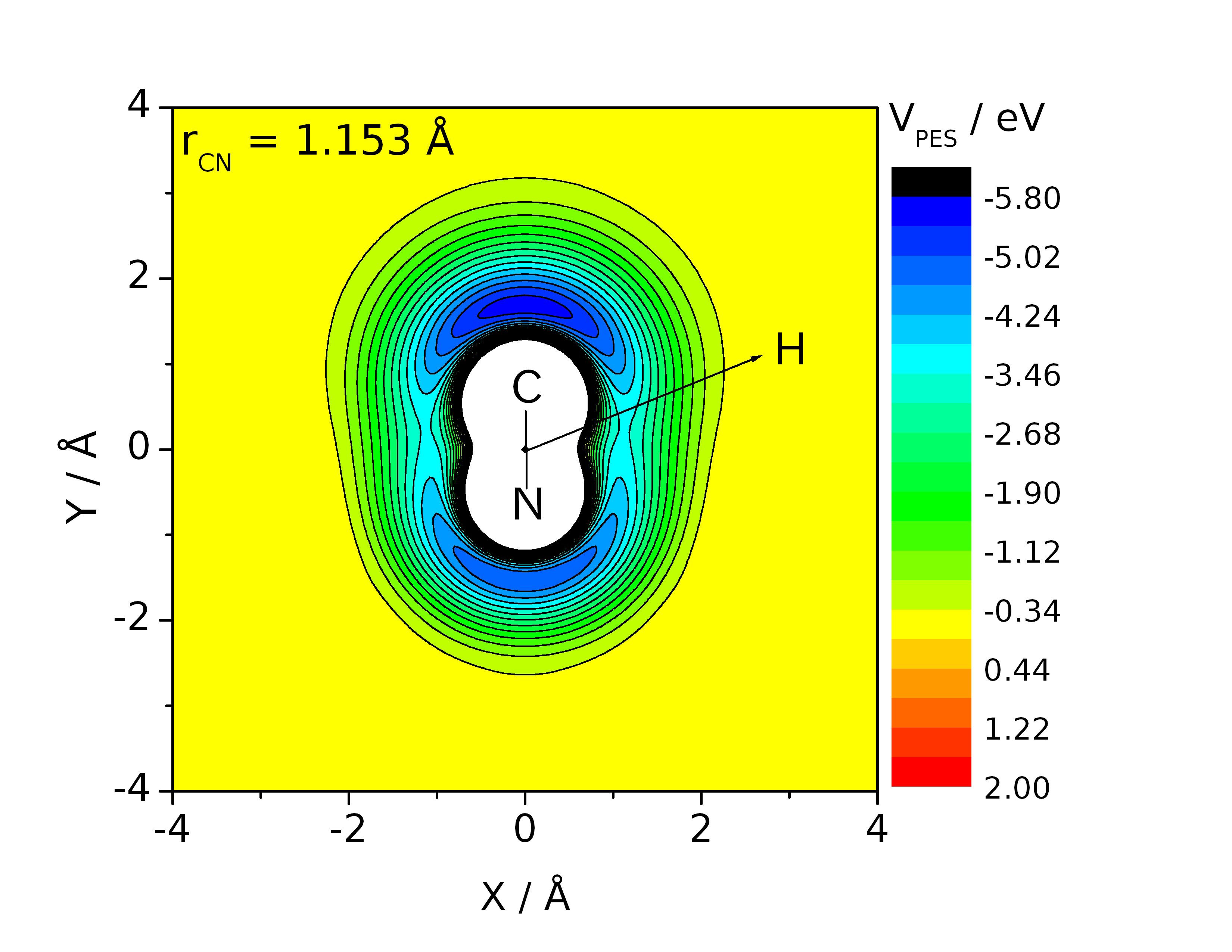}
\caption{Section of the PES of HCN at a fix $r_{\text{CN}}$ distance}
\label{fig:PES_2D}
\end{figure}

There is another numerical issue regarding the very long trajectories at low collision energies which makes impossible the evaluation of $P_r$ in Eq.~(\ref{Pr_gen_long}-\ref{Pr_neut_long}). Based on the Nyquist condition, the maximum frequency that can be represented by Fourier transformation is $\omega_{\text{top}} \propto n_{\text{FFT}} / t_{\text{traj}}$, where $n_{\text{FFT}}$ is the number of sampling points and $t_{\text{traj}}$ is the integration time of the trajectory. This means that for very long trajectories, especially at low collision energies, the cut-off ($\omega_{\text{top}}$) might be too small to represent properly the low frequency part of the spectrum which has significant contribution to the radiative probability.  Based on the spectra of individual collision (see Fig~\ref{fig:RA3D_spectrum}), in general at least $\omega_{\text{top}}=10000$ cm$^{-1}$ is required in this reaction for the frequency integral in Eq.~(\ref{Pr_gen_long}-\ref{Pr_neut_long}). In principle, we could increase $n_{\text{FFT}}$ to expand the cut-off frequency up to $\omega_{\text{top}}=10000$ cm$^{-1}$, however,  it would not be feasible anymore to evaluate the Fourier transformation within reasonable computation time for all the trajectories.  
Although we are not able to calculate $P_r$ for such trajectories, we cannot discard them, since these collisions would have a considerable contribution to the cross section. To take into account such trajectories efficiently, we introduced the above-mentioned double-layered MC scheme.
Besides the integration of the equations of motion, the Fourier transformation is the other  computationally intensive numerical procedure in the calculation of radiative association. It is not useful to apply the same number of FFT sampling point for each trajectory, since the time length of the collisions can be quite diverse and we do not need the spectrum at too high frequencies. To make more efficient the evaluation of $P_r$ we dynamically adjusted $n_{\text{FFT}}$ for each trajectory based on their integration time to make possible the representation of the emission spectrum at least up to $\omega_{\text{top}}=10000$ cm$^{-1}$. The $P_r$ of too long trajectories were not evaluated, where we could not achieve at least $\omega_{\text{top}}=10000$ cm$^{-1}$ with maximum $n_{\text{FFT}}=2^{15}$. Such trajectories are automatically considered as reactive events in Eq.~(\ref{charfunc}), owing to their enormous radiative power.

\subsubsection{Monte Carlo evaluation of the cross section}

Owing to the aforementioned issue with the long complex-forming trajectories, the double-layered MC method is used to obtain reliable cross sections. In this case of the too long trajectories, where the cut-off frequency makes impossible to obtain the radiative probability, we automatically consider such trajectories as reactive events in Eq.~(\ref{Improved_MC_cross}) without calculating $P_r$. The double-layered MC method requires the fictive photon number of the field, $\Omega$  as an input parameter in Eq.~(\ref{Improved_MC_cross}). The photon number $\Omega$ used in the calculation, should be small enough to avoid saturation of radiation, which in turn results in an underestimation of $P_r$ through the division by $(\Omega+1)$ in Eq.~(\ref{Improved_MC_cross}). As a rule of thumb, we recommend that $\Omega$
should be chosen to produce less reactive trajectories than 10\%-15\% of $N_\text{tot}$

\section{\label{sec:level4} Global potential energy and dipole surface of H + CN system}
\subsection{\label{sec:level5} Details of the quantum chemical calculations}
The Molpro package \citep[]{Molpro}
% Molpro seems to be missing in bib.
was used for the quantum chemical calculations. The potential energy and the dipole moment surface are calculated with the explicit correlated internally contracted multireference configuration interaction method (icMRCI-F12) using the aug-cc-pVTZ-F12 basis set as implemented in Molpro. All calculations were carried out in the $C_{s}$ point group. The reference wavefunction is obtained by using the state averaged complete active space self-consistent field (CASSCF) method with a full valance active space, and with six states (3$A'$ and 3$A''$) in the state average. The dipole moment of the system is calculated as an expectation value of the dipole operator.

In order to sample the configuration space of the system, we employed Jacobi coordinates of the HCN molecule denoted by ($r_{\text{CN}}$, $R$, $\theta$). We used 19 points for $r_{\text{CN}} \in [0.8,3.0] $ \AA, 46 points for $R \in [0.5,12.0]$ \AA, and 25 points for $\theta \in [0^{\circ},180^{\circ}]$ to represent the configuration space.

\subsection{\label{sec:level6}Global potential energy surface}
The calculated ab inito energies are interpolated by a 3D B-spline using the  BSpline package \citep{BSpline}. The obtained PES is attractive, there is no barrier in the entrance channel (see Fig.~\ref{fig:PES_2D}). Over the range of the 3D spline fitting we employed the following extrapolation formulas 
\begin{equation}
    V_{\text{sr}}(r_{\text{CN}},\theta,R)=a e^{-b R},
\end{equation}
which describes the short range part of the PES, and
\begin{equation}
    V_{\text{lr}}(r_{\text{CN}},\theta, R)=D_0 - C_6/R^6
\end{equation}
for the long-range extrapolation similarly as did by Ayouz et al.\cite{H3neg_Dipsurf}. The coefficients, $a$, $b$, $D_0$, $C_6$, are evaluated at each ($r_{\text{CN}}, \theta$) value. The short range extrapolation parameters, $a$ and $b$, are determined from two points: $V(r_{\text{CN}}, \theta, R=0.50\text{\AA})$ and $V(r_{\text{CN}}, \theta, R=0.52\text{\AA})$ at a given $R$, while the long-range parameters, $D_0$ and $C_6$  are obtained from $V(r_{\text{CN}}, \theta, R=11.00\text{\AA})$ and $V(r_{\text{CN}}, \theta, R=11.80\text{\AA})$ of the 3D spline potential values. 
\begin{figure}
\includegraphics[width=8.5cm,angle=0]{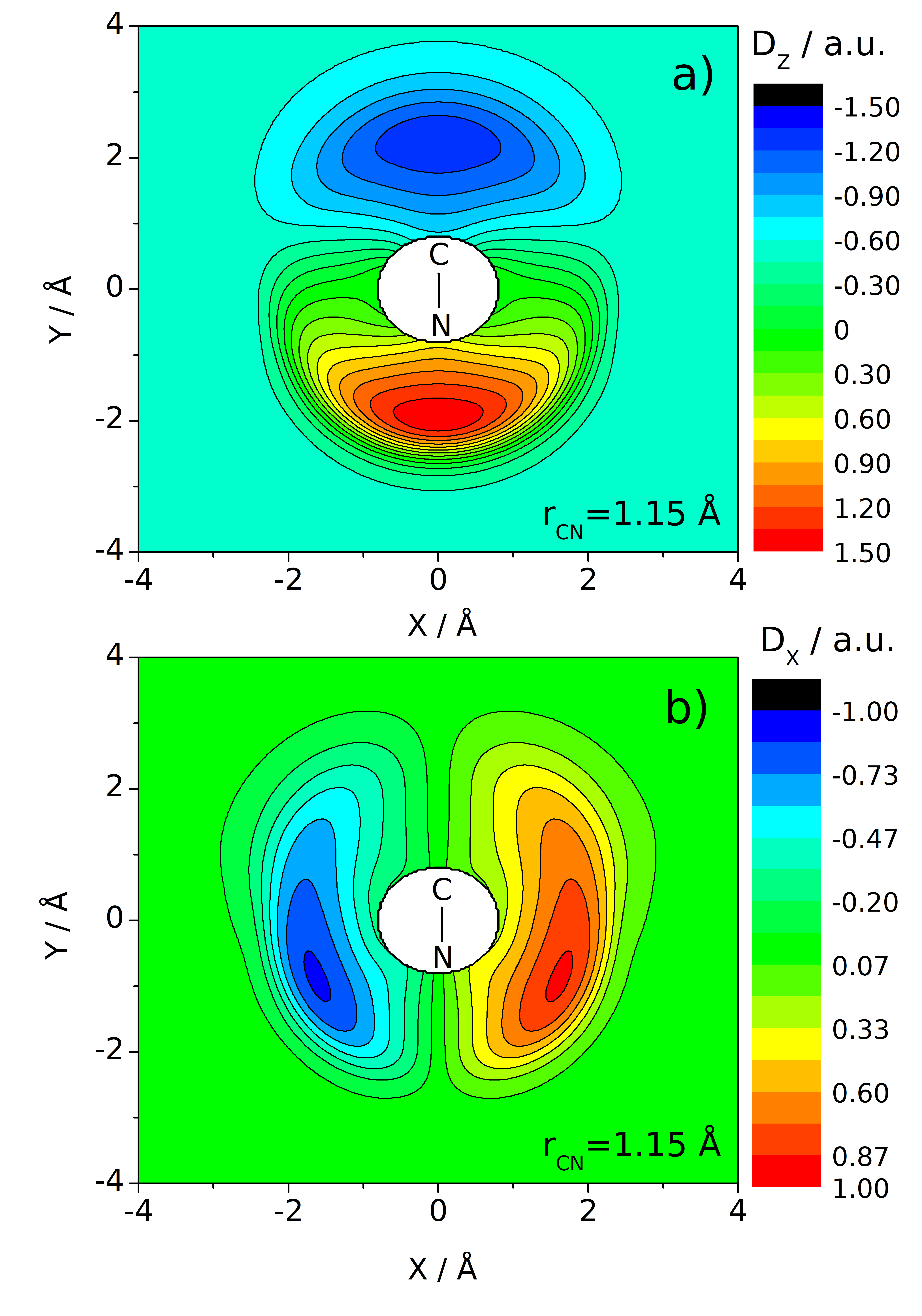}
\caption{Section of dipole surface of the HCN system. Panel a) shows z-component and panel b) displays the x-component of the dipole}
\label{fig:Dip_2D}
\end{figure}
In order to test the quality of the PES, we used the DVR3D code \cite{DVR3D} to compute the ro-vibrational quantum states. The results of these calculations are compared to the ro-vibrational energies obtained on the PES of Mourik et al.\cite{HCN_PES_Mourik} which is developed at CCSD(T) level for bound state calculations. In general, the ro-vibrational energies of our PES slightly underestimate (by 10-15 cm$^{-1}$)  the corresponding quantum states on the PES of Mourik et al. \cite{HCN_PES_Mourik} (see Fig. S1 in the Supplemental Material). Such deviation is acceptable for global PESs  which are used for collision dynamics and not solely for bound state computations in order to describe the ro-vibrational spectroscopy with high accuracy. 
% However, in contrast to the PES of Mourik et al., the long range part of the PES is represented properly is necessary in the accurate description of the collision dynamics.

\subsection{\label{sec:level7}Global dipole surface}
%One of the biggest shortcoming of polyatomic full dimensional RA calculations is the missing global dipole surfaces. Most of the dipole surfaces are developed merely for the calculation of spectral line intensities. 

Besides the PES, we also calculated and fitted the global dipole surface of the HCN system which is displayed in Fig~\ref{fig:Dip_2D}. 
The dipole vector is defined in the frame of principal axis of inertia  (PAI) as in Molpro. The $z$ component of the dipole is parallel with the CN bond
and the $x$ component is perpendicular to the CN bond.
For the fitting procedure, we used the following function forms for the two PAI component of the dipole ($D_y = 0$)
\begin{equation}
D_{x}= \sum_{i=0}^{3}\sum_{j=0}^{6}\sum_{k=1}^{6} C_{ijk} \, r_{\rm CN}^{i}\,\, e^{-A_{j} \left(R-B_{j}\right)^{2}} P_{k}^{1}\left(\cos \theta\right),
\end{equation}

\begin{equation}
D_{z}= \sum_{i=0}^{3}\sum_{j=0}^{6}\sum_{k=0}^{6} C_{ijk} \, r_{\rm CN}^{i}\,\, e^{-A_{j} \left(R-B_{j}\right)^{2}} P_{k}^{0}\left(\cos \theta\right),
\end{equation}
where $P_{k}^{m}\left(\cos \theta\right)$ factors are the associated Legendre polynomials and $A_j$, $B_j$, $C_{ijk}$ are constant parameters which are determined by nonlinear fitting. At long ranges ($R > 5.5$ \AA) the dipole surface is extrapolated with the dipole of the free CN diatom.
The comparison of the fitted dipole surface of the present work and that of Mourik et al.\cite{HCN_PES_Mourik} can be found in Fig. S2 in the Supplemental Material.  Around the equilibrium structures the two surfaces show a good agreement. However, the dipole surface of  Mourik et al.\cite{HCN_PES_Mourik} is represented only in a limited region of the configuration space that is useful for the calculation of spectral line intensities of a bound system, but not for the description of a collision processes.

%\begin{figure}
%\includegraphics[width=8.0cm,angle=0]{figs/Dipole_comparison.jpg}
%\caption{Dipole of H + CN dipole energy surface}
%\label{fig:Dip_comp}
%\end{figure}

\section{\label{sec:level8} Results and discussion}
\subsection{\label{sec:level9} Reduced dimensional QCT and quantum dynamics calculations}
Besides the full dimensional QCT computations, we also performed reduced dimensional calculations in 1D by using our semiclassical QCT method and quantum mechanical perturbation theory (QM). (See Appendix for the computational details of QM.) For these calculations we utilized the 1D sections of potential energy and dipole surfaces along the collinear H-CN and H-NC axis. Such reduced dimensional models may help us to test the performance of classical/semiclassical methods when full dimensional quantum dynamical calculations are not available \citep{QCT_reduced_1,QCT_reduced_2}.

\begin{figure}
\includegraphics[width=8.5cm,angle=0]{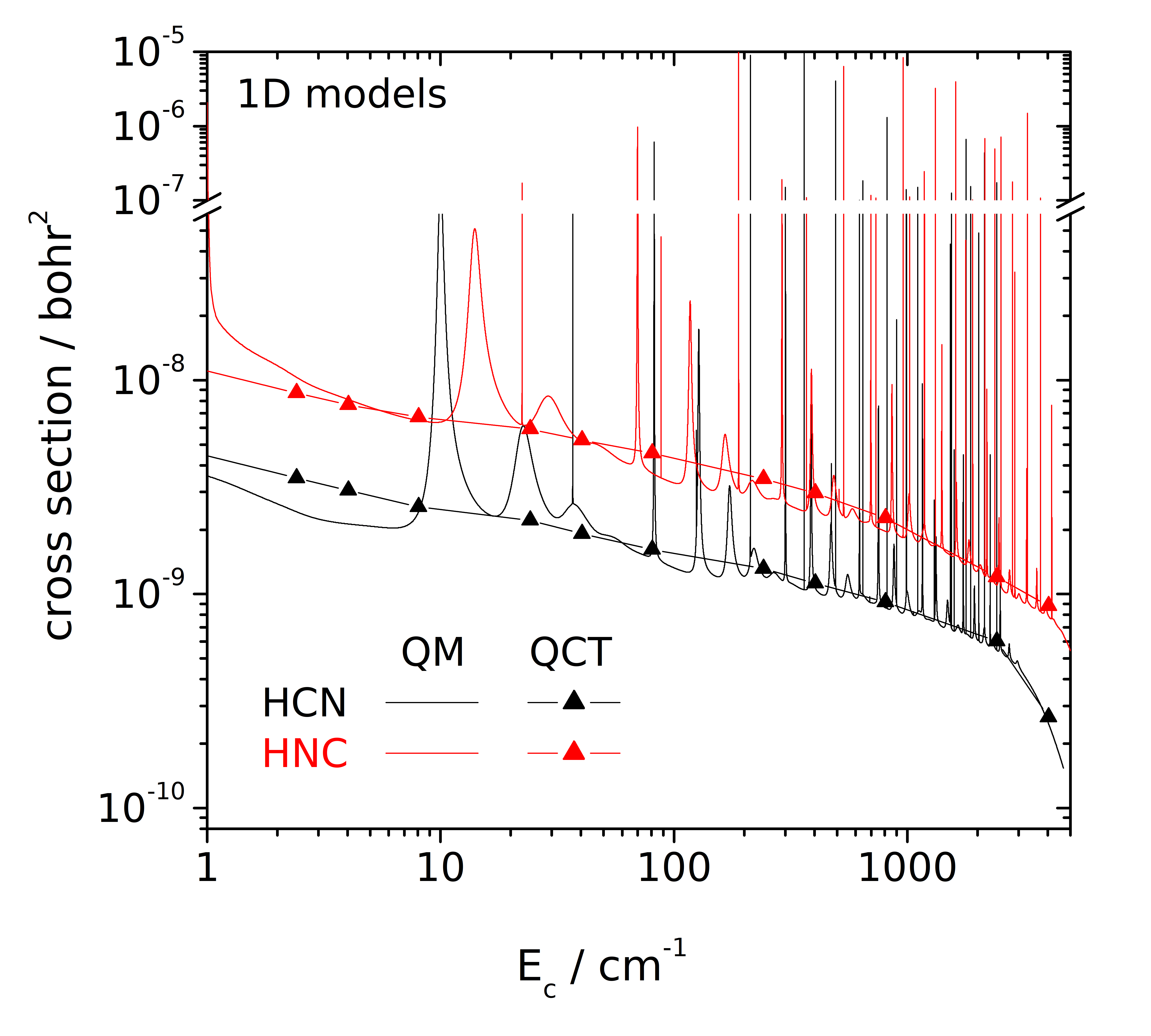}
\caption{Reduced dimensional (1D) cross section of the HCN and HNC formation in  H + CN collisions obtained by quantum mechanical perturbation theory (QM) and the QCT method}
\label{fig:1D_cross}
\end{figure}

Fig.~\ref{fig:1D_cross} shows the cross sections for the formation of HCN and HNC obtained by 1D reduced dimensional QCT and QM calculations. In complex-forming collision when the tunneling is negligible, then the QCT method can remarkably well reproduce the baseline of the quantum dynamical cross section. Similar good agreement is observed for diatomic molecule formation when the capture is not hampered by a potential barrier in the entrance channel.

Based on this, we reckon here that the full dimensional QCT calculations can provide reliable cross section for a wide range of collision energies apart from the resonances.

\subsection{\label{sec:level10} Full dimensional QCT calculations}

\begin{figure}
\includegraphics[width=8.5cm,angle=0]{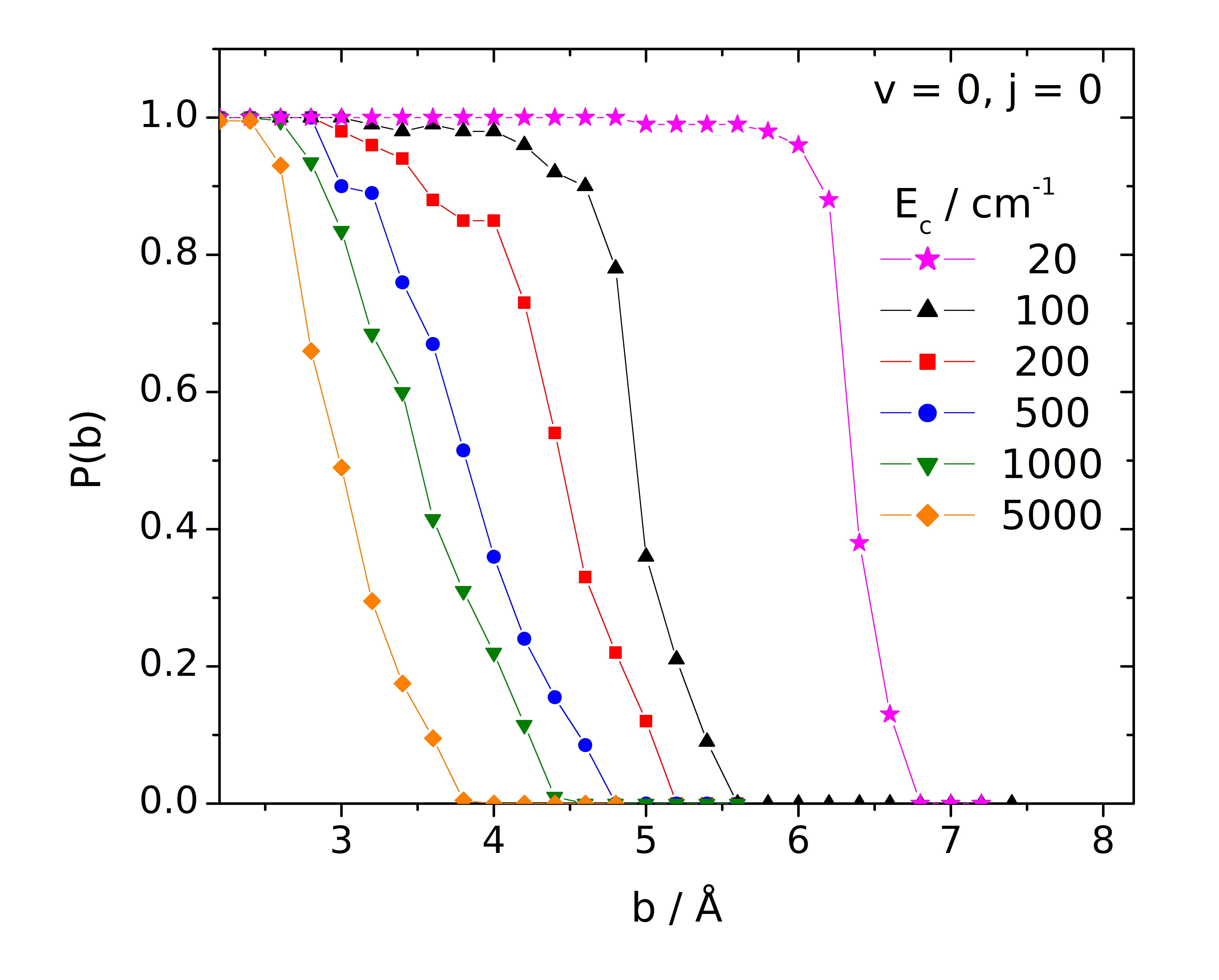}
\caption{Probability of complex formation (capture) as a function of impact parameter in  H + CN collisions in full dimension.}
\label{fig:opacity}
\end{figure}

The full dimensional description of RA process is highly desired, since there are several vibrational degrees of freedom of polyatomic molecules that may contribute to the radiation process. QCT provides an efficient way to calculate the detailed dynamics of molecule formation through RA in arbitrary dimension when the global PES and dipoles surface is available. 

The RA process is efficient when the system fall into the deep potential well, and the reactants spend several ro-vibrational periods together in the collision complex. That is why we calculated the radiative power only for complex-forming trajectories where the change of the dipole is considerable during the collision. The complex is defined by a geometric and energetic constraint: the center of mass of CN diatom and the hydrogen atom has to be closer than 4.5 \AA,  and the potential energy has to reach at least the half of HCN potential well, -2.6 eV during the collision process (see Fig.~\ref{fig:PESscheme}). Based on this definition, the non complex-forming trajectories has negligible contribution to the radiated probability. 

The opacity functions of complex-formation is displayed in Fig.~\ref{fig:opacity}. The probability of capture is almost unity even at large impact parameters when the collision energy is low. This is a clear sign of the long-range attraction, which makes the capture more efficient under cold interstellar circumstances. The corresponding cross sections  for complex-formation (see in Fig S3. in the Supplemental Material) shows a divergent cross section as the collision energy diminishes.

\begin{figure}
\includegraphics[width=8.5cm,angle=0]{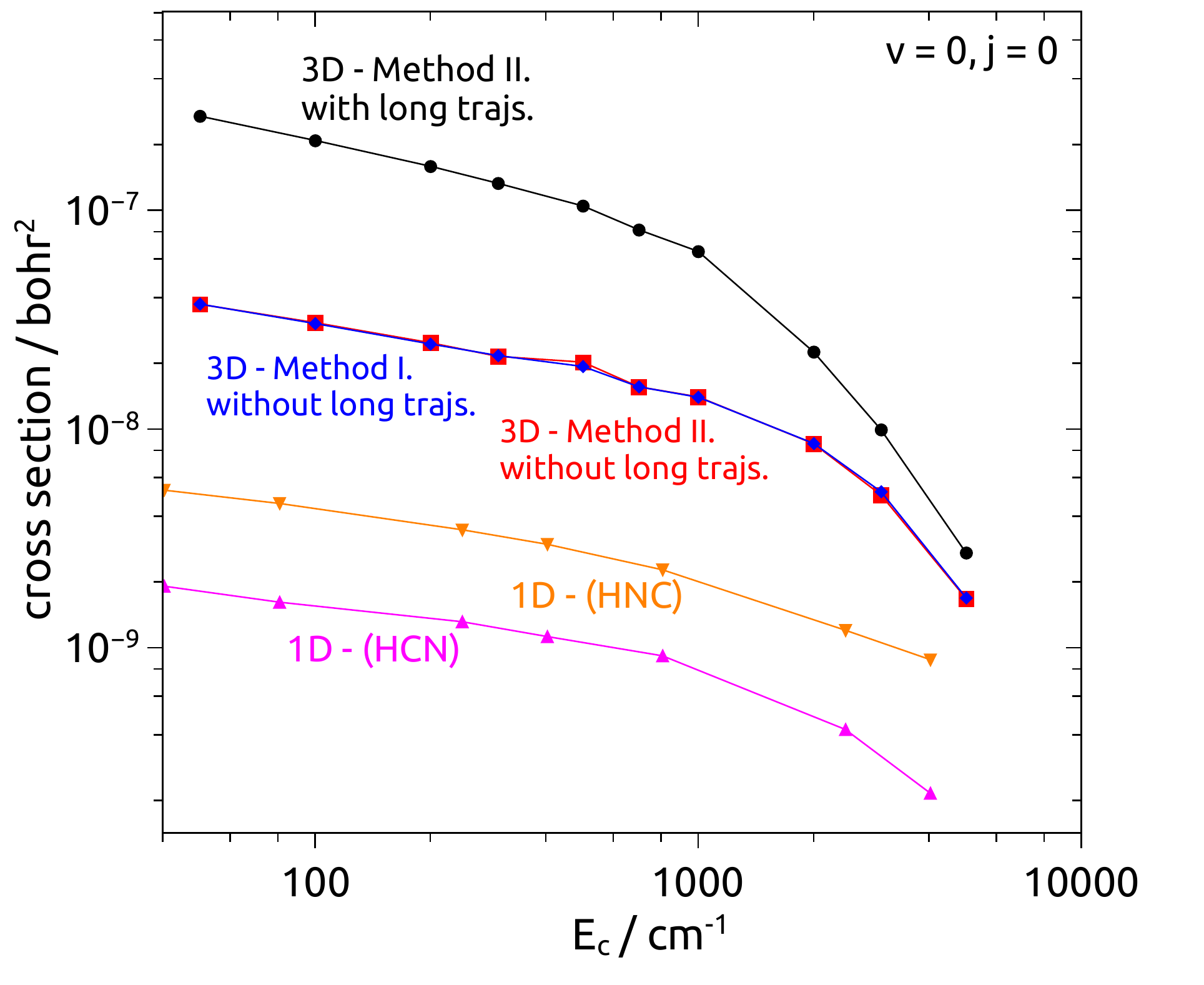}
\caption{Comparison of the cross sections obtained by different methods for the formation of HCN/HNC through radiative association.}
\label{fig:3Dcrosses}
\end{figure}

The full dimensional RA cross section for the collision of CN($v=0,\,j=0$) and hydrogen atom is shown in Fig.~\ref{fig:3Dcrosses}. It has a divergent character at low collision energies similarly to the capture cross sections, since at low collision energies the reactants spend more time together. Therefore, the probability of photon emission in the RA process is much higher at cold temperatures. Comparison of the two Monte Carlo schemes is also presented in Fig.~\ref{fig:3Dcrosses}. In Method I. the radiation is treated as a classical process where the trajectories with too long lifetime are discarded because of the above-mentioned numerical issues, while in Method II. such trajectories are included due to the second Monte Carlo step which mimics the discrete nature of photon emission. The outcome of Method I. and II. clearly shows that the two different Monte Carlo schemes provide the same result when the extreme long trajectories are discarded. Nevertheless, Method II. allows us to take into account the trajectories with long lifetime that have a considerable contribution to the reactivity. The cross section is larger with almost one order of magnitude at low and intermediate collision energies than those of obtained by omitting the trajectories with extreme long lifetime.

In Fig.~\ref{fig:3Dcrosses} we compare the outcome of the reduced dimensional (1D) and the full dimensional QCT calculations as well. As can be seen, the full dimensional calculations provide significantly larger cross section than those of the reduced dimensional model. The difference is almost 2 orders of magnitude in a wide range of collision energies. This means that the C-N stretching and H-C-N and/or H-N-C bending modes of HCN/HNC molecule (as well as the large amplitude motion of the hydrogen atom around the CN molecule in the deep potential well) --~which were kept frozen in the 1D model -- have remarkable impact on the radiative probability and molecule formation. Moreover, if we assume similar performance from the QCT method in 1D and full dimension (the QCT method can reproduce the baseline of the QM cross section), then the full dimensional QCT calculations provide a lower bound to the exact quantum mechanical cross section owing to the lacking resonances. 

It has to be stressed that our QCT calculations cannot distinguish the formation of HCN and HNC isomers by radiative association. However, as we will see below, this is unnecessary based on the analysis of emission spectra.

\subsection{\label{sec:EmissionSpectra} Emission spectra and the mechanism of radiative association in H + CN collisions}

Besides the cross section, our method also allows us to extract the spectrum of the emission in collisions. By using the Wiener-Khinchin theorem, it can be shown that this spectrum is similar to that is obtained from dipole-dipole auto-correlation function, which is usually used to calculate vibrational spectrum from molecular dynamics simulations\citep{QCT_6,QCT_7,QCT_8}. However, in our case, we calculate a transient (collision induced) vibrational emission spectrum in contrast to the infrared absorption spectrum of stable molecules. This vibrational information of the transient collision complex may shed light on the mechanism of radiative association and how it is depends on the collision energy and the initial ro-vibrational state of reactants.
\begin{figure}
\includegraphics[width=8.5cm,angle=0]{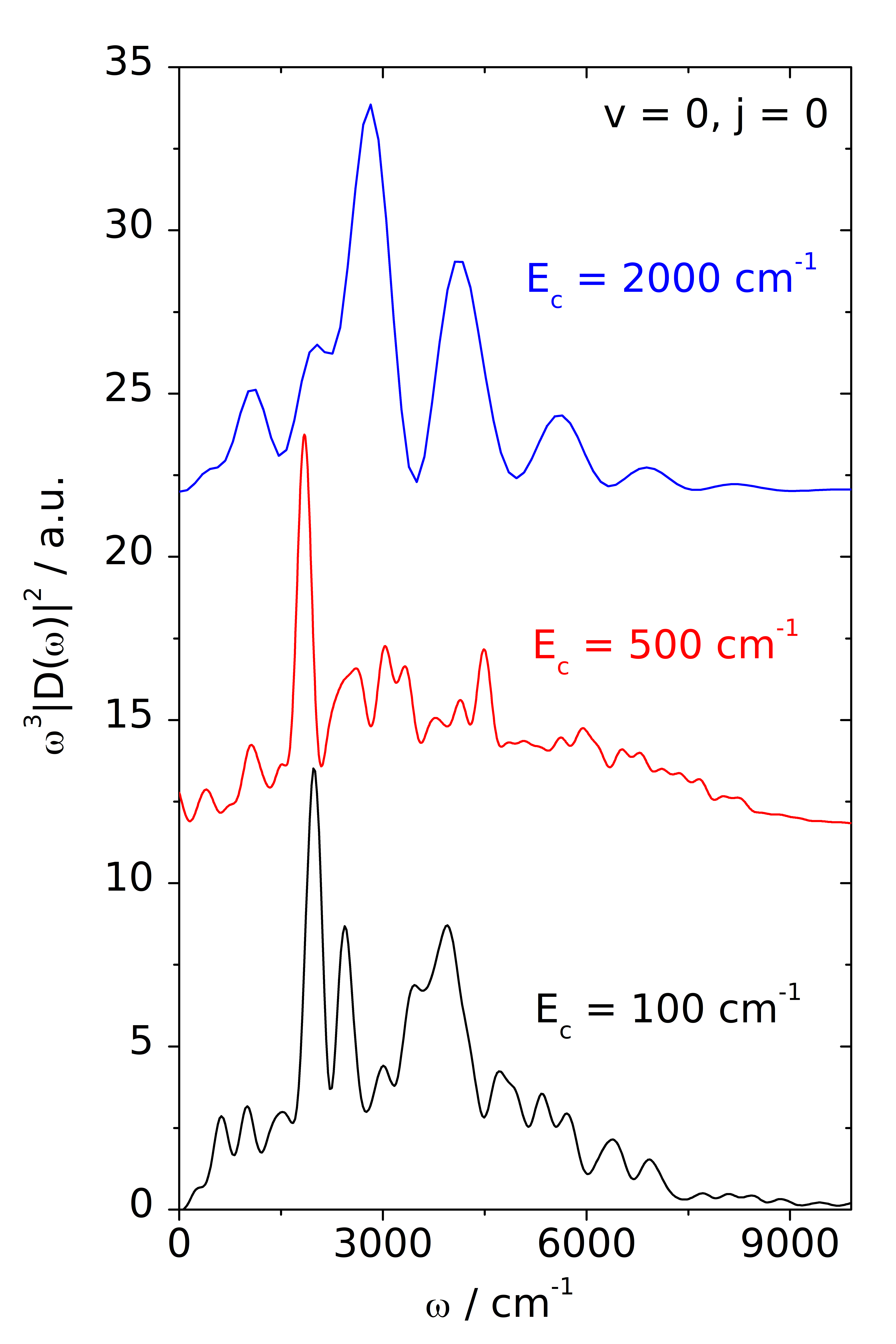}
\caption{Normalized spectrum of the radiation for some representative trajectories in  H + CN collisions at three different collision energies.}
\label{fig:RA3D_spectrum}
\end{figure}

In Fig.~\ref{fig:RA3D_spectrum} the emission spectrum of representative collisions at three different energies are displayed. Unfortunately, we could not calculate a reliable averaged spectrum for low collision energies because of the above-mentioned numerical issues with the long trajectories (Nyquist condition in the FFT procedure). Nevertheless, by analyzing a large number of individual trajectories, we may draw a conclusion about the mechanism of RA process in HCN/HNC system. At high collision energies the combination bands of the stretching with the bending modes give the biggest contribution to the radiative probability (see Fig.~\ref{fig:RA3D_spectrum} and Tab.~\ref{tab:freq}) which is a clear sign of the strong anharmonicity caused by the high-energy content of the system. This is not surprising, since the timescale of the fast collisions is comparable to the fastest degree of freedom of the system. In most of the energetic collisions, there is no time for several periods of slow motions as the large amplitude oscillation of H around the CN molecule or the pure C-N stretching. However, when the collision energy is small enough, then efficient pathways are opened for energy-flow through the slow degrees of freedom, the pure harmonic frequencies will appear in the spectrum. Hence, for instance below $E_c$~=~500 cm$^{-1}$ the harmonic C-H, N-H and C-N stretching modes are dominating (see Fig.~\ref{fig:RA3D_spectrum}). Furthermore, in many cases, instead of the bending modes, a large amplitude motion (LAM) can be observed, when the H atom orbits the CN diatom. This LAM can be considerably efficient for RA since the dipole of system is changing extensively and swiftly.
\begin{table}
\caption{Harmonic vibrational frequencies of the HCN and HNC molecules on the PES of this work and their first overtones and combinations. \\}     
\label{tab:freq}
\begin{tabular}{ll|ll}
\hline
\hline
\multicolumn{2}{c}{HCN} & \multicolumn{2}{c}{HNC} \\
\hline
~Mode   & $\nu$(cm$^{-1}$)~ & ~Mode & $\nu$(cm$^{-1}$)~  \\ 
\hline
~~$\nu_{\text{CH}}~~~$    & ~3310~  & ~~$\nu_{\text{NH}}$~~~     & ~3614~\\ 
~~$\nu_{\text{CN}}~~~$    & ~2081~  & ~~$\nu_{\text{CN}}$~~~     & ~2012~\\ 
~~$\nu_{\text{bend}}~~~$  & ~704~ & ~~$\nu_{\text{bend}}$~~~    & ~462~\\ 
                      &                                 &       \\
~~$2\nu_{\text{CH}}$~~~   & ~6620~ & ~~$2\nu_{\text{CH}}$~~~    & ~7228~\\ 
~~$2\nu_{\text{CN}}$~~~   & ~4162~ & ~~$2\nu_{\text{CN}}$~~~    & ~4024~\\ 
~~$2\nu_{\text{bend}}$~~~ & ~1408~  & ~~$2\nu_{\text{bend}}$~~~  & ~924~\\ 
                      &                                 &       \\
~~$\nu_{\text{CH}}+\nu_{\text{CN}}$~~~   & ~5391~ & ~~$\nu_{\text{NH}}+\nu_{\text{CN}}$~~~    & ~4076~\\ 
~~$\nu_{\text{CH}}+\nu_{\text{bend}}$~~~   & ~4014~ &  ~~$\nu_{\text{NH}}+\nu_{\text{bend}}$~~~   & ~5626~\\ 
~~$\nu_{\text{CN}}+\nu_{\text{bend}}$~~~   & ~2785~ &  ~~$\nu_{\text{CN}}+\nu_{\text{bend}}$~~~   & ~2474~ \\
\hline
\hline
\end{tabular}
\end{table}

We may make a further important observation by realizing that the radiative power has significant contribution up to $\omega_{\text{top}}$~=~10000 cm$^{-1}$ in all the collisions (see Fig.~\ref{fig:RA3D_spectrum}). This means that the emitted photon takes away maximum $\hbar \omega_{\text{top}}$ energy, and the formed molecule has  a maximum binding energy $E^{\text{max}}_{\text{bind}}\approx \hbar \omega_{\text{top}} - E_c$. Hence, the formed HCN/HNC system is just below its dissociation limit, with energy content far above the barrier separating the HCN and HNC molecules.
Based on this, it does not make sense to distinguish the formation of HCN and HNC molecules in the present QCT calculations. In the interstellar medium where the three-body collision are negligible, the hot HCN/HNC system formed by RA can stabilize either in HCN or HNC molecule by the emission of a second photon.

%============================================================================================================
\section{Conclusion}
In this work we have presented an efficient semiclassical method to calculate the detailed dynamics for the formation of arbitrary size molecule by radiative association in absence of electronic transitions. In order to test our method, we also developed a global full dimensional potential and dipole surface for the H + CN nonreactive collisions. The PES and dipole surface are available as Supplementary Material. Our method is a combination of the QCT method and the classical Larmor formula, which provides the radiative power in collisions. Owing to the QCT method, we may calculate ro-vibrationally quantum state resolved cross sections of RA. We also analyzed the numerical issues regarding the RA calculation using the QCT method, and provided solution for them. To this end we also developed in this work a new efficient double-layered Monte Carlo algorithm which can mimic the quantized nature of the emitted photon. Based on our calculation, we have shown that the full dimensional treatment of the RA process is important, in contrast to some previous works where the collision of polyatomic molecules are described solely in 1D. In full dimension, the RA cross section is almost 100 times larger in a wide range of collision energies than that of the reduced dimensional (1D collinear) model. Furthermore, our method also allows us to obtain the emission spectrum of RA process. Analysis of these transient emission spectra can provide the details of the RA mechanism of the molecule in question.

\section{Appendix}

\subsection{\label{sec: App-Emax} The upper limit of the frequency integral and the minimum energy ro-vibrational state of molecules formed by radiative association}
\begin{figure}
\includegraphics[width=8.5cm,angle=0]{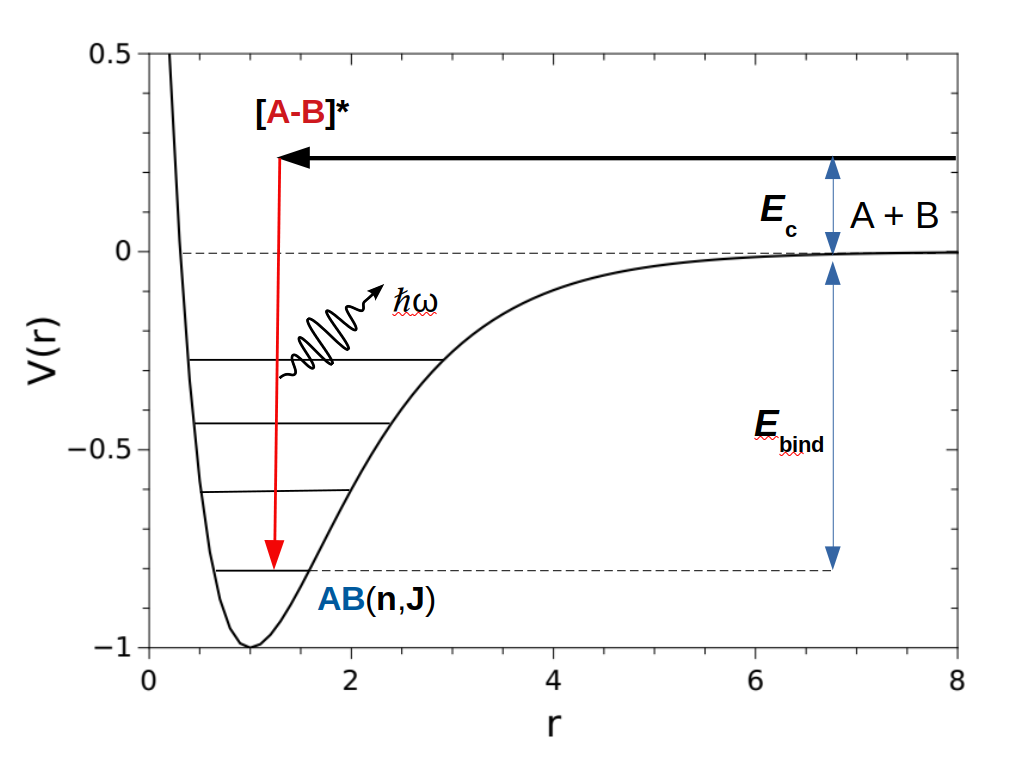}
\caption{Schematic representation of the radiative association process for the formation of a hypothetical AB($\mathbf{n}$, $\mathbf{J}$) molecule in a given ro-vibrational quantum state characterized by a collection of vibrational quantum numbers, $\mathbf{n}$, and angular momentum, $\mathbf{J}$.}
\label{fig:scheme}
\end{figure}

The upper limit of the frequency integral, $\omega_\text{max}$  in Eqs.~(\ref{Pr_gen_simple},\ref{Pr_neut_simple},\ref{Pr_gen_long} and \ref{Pr_neut_long}) requires the determining the maximum binding energy (or the minimum energy content) of molecules formed after the emission of a photon by considering the conservation of the total angular momentum (see Fig.~\ref{fig:scheme})
\begin{equation}
    \hbar \omega_\text{max} = E_c + \text{max}\left[E_{\text{bind}}(\mathbf{J})\right], 
    \label{eq:Emax}
\end{equation}
where $E_c$ is the collision energy and $E_{\text{bind}}(\mathbf{J})$ is the binding energy of the formed molecule at a given total angular momentum, $\mathbf{J}$, as displayed in Fig.~\ref{fig:scheme}.
% Peter: Consider changing this definition of binding energy so that it is a positive number for a bound system. Then it would be consistent with the last paragraph of section VC. If you agree, then also E_0^eq in Eq. (30) should probably also be defined to be a positive number.

When the potential energy is deep enough, like in the case of the HCN/HNC molecule, then the value of $\omega_\text{max}$ can be chosen smaller than that demanded by the possible maximum binding energy of the molecule with a given angular momentum. Such simplification is possible since for a general molecule the C-H, O-H and N-H stretching are the largest frequency harmonic modes (see e.g. Tab.~\ref{tab:freq}). Furthermore, the overtones and the combination bands of these largest and other lower frequency harmonic modes can give a considerable radiative contribution up to $\omega_{\text{top}} \approx$ 10000-15000 cm$^{-1}$ for a general molecule. Therefore, we may approximate the upper limit of the frequency integral in Eqs.~(\ref{Pr_gen_simple},\ref{Pr_neut_simple},\ref{Pr_gen_long} and \ref{Pr_neut_long}) as $\omega_{\text{max}}=\omega_{\text{top}} $. However, when the potential well of the molecule is shallow (less than 15000 cm$^{-1}$ deep), we have to optimize $E_{\text{bind}}(\mathbf{J})$ at constant $\mathbf{J}$ values. In the remainder of this section, we give a practical solution how to estimate the maximum of $E_{\text{bind}}(\mathbf{J})$ which is changes from collision to collision, owing to the varying total angular momentum in different trajectories.

First, we assume that the total internal energy of the molecule -- measured from the bottom of the potential well -- is separable: $E_{\text{mol}}=E_{\text{vib}}+E_{\text{rot}}$, where
\begin{equation}
E_{\text{vib}}=\sum_{i=1}^{3N-6}\left(\frac{1}{2}+n_i\right)\hbar\nu_{i}
\label{Evib}
\end{equation}
is assumed to be the sum of quantum mechanical harmonic mode energies, where $n_i$ is the vibrational quantum number and $\nu_i$ is the frequency of the $i$th mode. The rotational energy can be approximated as the energy of a classical 3D rigid-rotor (RR) 
\begin{equation}
E_{\text{rot}}=\frac{1}{2}\mathbf{J}^{\rm T}\mathbf{I}^{-1}\mathbf{J},
\label{Erot}
\end{equation}
where $\mathbf{J}$ is the total angular momentum of the formed molecules in the A + B $\rightarrow$ AB + $\hbar\omega$ collisions, and $\mathbf{I}$ is the tensor of inertia of the AB molecule. By considering that the emitted photon takes away 1$\hbar$ angular momentum, the total angular momentum of the formed molecules in the A + B $\rightarrow$ AB + $\hbar\omega$ collisions is
\begin{equation}
\mathbf{J}=\mathbf{L}_{\text{orb}}+\mathbf{J}_{\text{A}}+\mathbf{J}_{\text{B}}-\hbar\mathbf{E}
\label{Jtot}
\end{equation}
where $\mathbf{L}_{\text{orb}}$ is the orbital angular momentum, $\mathbf{J}_{A}$ and $\mathbf{J}_{B}$ are the rotational angular momenta of the fragment A and B. Furthermore, $\mathbf{E}$ is a unit vector that determines the direction of the angular momentum of the emitted photon. The rotational quantum numbers, $j_{A}$ and $j_{B}$ determine the magnitude of $|\mathbf{J}_{A}|=\hbar\sqrt{j_{A}\left(j_{A}+1\right)}$ and $|\mathbf{J}_{B}|=\hbar\sqrt{j_{B}\left(j_{B}+1\right)}$, while their orientation is sampled randomly, which means it is different for each trajectory. The impact parameter, $b$ and collision energy determine the magnitude of the orbital angular momentum
\begin{equation}
    |\mathbf{L}_{\text{orb}}|=b\sqrt{2\mu E_{c}},
\end{equation}
while its well-defined orientation depends on the simulation setup. Moreover, the orientation of the angular momentum, $\mathbf{E}$, taken by the photon is also supposed to be sampled randomly. Based on these considerations, the possible energy content of the molecules formed in RA process depends on the initial conditions of the reactants, owing to the varying orientation of the angular momenta. In order to find $\omega_{\text{max}}$, we have to minimize $E_{\text{mol}}=E_{\text{vib}}+E_{\text{rot}}$ at certain $\mathbf{J}$ values. By using the harmonic oscillator (HO) approximation in Eq.~(\ref{Evib}), we have to minimize $E_{\text{mol}}$ along the normal mode coordinates. Hence, by optimizing the value of vibrational quantum numbers, $(n_1,n_2,\ldots,n_{3N-6})$ as well as the rotational energy along the normal coordinates at each fixed set of vibrational quantum numbers, we obtain the maximum frequency as
\begin{equation}
  \hbar \omega_{\text{max}} = E_c  + E_0^{\text{eq}} - \text{min}\left[E_{\text{vib}}+E_{\text{rot}}\right],
\end{equation}
where $E_0^{\text{eq}}$ is the depth, to the global minimum, of the potential well. Furthermore, the amplitude of the normal modes is determined by the vibrational quantum numbers. This procedure can provide the exact $\omega_{\text{max}}$ within the HO-RR approximation. This minimization is supposed to be done for every trajectory, which can be computationally expensive for bigger molecules.

In order to provide a simpler estimation of $\omega_{\text{max}}$, we assume that the minimum of $E_{\text{mol}}$ is at the zero-point vibration level of the lowest energy isomer of the formed molecule (corresponding to the absolute minimum of the PES) regardless of the rotational energy. Furthermore, to avoid the minimization along the normal coordinates, we may assume that the minimum of the rotational energy (at a given $\mathbf{J}$) is at the equilibrium structure of the molecule. Hence, the maximum frequency of the emitted photon in a polyatomic RA process can be estimated as
\begin{equation}
  \hbar \omega_{\text{max}} \approx E_c + E_0^{\text{eq}} - \frac{1}{2}\sum_{i=1}^{3N-6} \hbar \nu_{i} - \frac{1}{2}\mathbf{J}^{\rm T}\mathbf{I}_{\text{eq}}^{-1}\mathbf{J},
\end{equation}
where $\mathbf{I}_{\text{eq}}$ is the tensor of inertia at the equilibrium geometry.

\subsection{\label{sec:App_Dip} Correction of the dipole for non-interacting reactants} 
When one wants to use Eq.~(\ref{Pr_neut_long}) to obtain the correct radiative power that results in molecule formation, then the radiation of the non-interaction reactants can be eliminated by using the induced dipole outside the interaction zone and the total dipole inside the zone. Thus, the corresponding dipole is 
\begin{equation}
\mathbf{D}\left(\mathbf{q},t\right)=\left\{ \begin{array}{ccc}
\mathbf{\widetilde{D}}\left(\mathbf{q},t\right)-\mathbf{\widetilde{D}}_0\left(\mathbf{q},0\right) & \:\text{if}\: & \mathbf{q}\notin\left[\text{int. zone}\right] \, \\
\\
\mathbf{\widetilde{D}}\left(\mathbf{q},t\right) & \:\text{if}\: & \mathbf{q}\in\left[\text{int. zone}\right],
\end{array}\right.
\label{interactionzone2}
\end{equation}
where $\mathbf{\widetilde{D}}_0(\mathbf{q},0)$ is the dipole of the non-interacting fragments at a given $\mathbf{q}$ molecular arrangement at time $t=0$. This is illustrated by non-zero value of $D_z$ for H far away in Fig.~\ref{fig:Dip_2D}a.
In the transition from inside to outside the interaction zone a switching function is applied as described for the alternative method in section \ref{sec:WhichDipole}.

\subsection{Quantum mechanical perturbation theory}
The quantum mechanical perturbation theory treatment of the radiative stabilization yields a Golden rule-like formula for the cross section \cite{BabR95,Rev_theo}

\begin{equation}
\sigma{\left(E_{c}\right)} = 
\frac{64 \pi^{5} f_{\text{stat}}}{3k_{\text{ini}}^{2} \, 4\pi\epsilon_{0} } 
\sum_{J,j',v'}  \frac{ S_{J,j'}} {\lambda^{3}_{J,j',v',E_{c}}}
{\left| M_{J,j',v',E_{c}} \right|^{2}},
\label{quant_cross}
\end{equation}

\noindent
where the sum runs over the initial angular momentum $(J)$ and final rotational $(j')$ and vibrational $(v')$ quantum numbers. $k_{\text{ini}}=\sqrt{\left(2 \mu E_{c} \right)}/\hbar$ is the wavenumber, where $\mu$ is the reduced mass of the colliding fragments, $S_{J,j'}$ is the H\"onl-London factor, $\lambda_{J,j',v',E_{c}}$ is the wavelength of the emitted photon. The transition dipole matrix elements are defined as
\begin{equation}
M_{J,j',v',E_{c}} =  \left\langle \varphi_{j',v'}^{\text{fin}}\left(r\right)\left|\hat{D}\left(r\right)\right|\chi_{J,E_{c}}^{\text{ini}}\left(r\right)\right\rangle,
\end{equation}

\noindent
where $\hat{D}(r)$ is the operator of the dipole moment function, $\chi_{J,E_{c}}^{\text{ini}}(r)$ is the radial part of the energy normalized scattering wavefunction of the initial state and $\varphi_{j',v'}^{\text{fin}}(r)$ is the radial part of the final ro-vibrational wavefunction, normalized to unity.

\begin{acknowledgments}
This article is based upon work from COST Action CA18212—Molecular Dynamics in the GAS phase (MD-GAS), supported by COST (European Cooperation in Science and Technology). Support   from   the   Kempe   Foundation   is   gratefully
acknowledged. M.G. acknowledges support from the Knut and
Alice Wallenberg Foundation. Computational resources provided by the Swedish National Infrastructure for Computing (SNIC) at HPC2N are acknowledged.
\end{acknowledgments}

\section*{Author Declarations}
\subsection*{Conflict of Interest}
The authors have no conflicts to disclose.

\section*{Data availability}
The data that support the findings of this study and the global potential and dipole surface are openly available as Supplemental Material.

\nocite{*}

\bibliography{HCN}% Produces the bibliography via BibTeX.

\end{document}